\documentstyle[preprint,aps,eqsecnum]{revtex}

\begin{document}
\draft
\title{{\Large {\bf Phase Transitions of the Flux Line Lattice in
High-Temperature Superconductors with Weak Columnar and Point Disorder}} }
\author{Yadin Y. Goldschmidt}
\address{Department of Physics and Astronomy\\
University of Pittsburgh\\
Pittsburgh, PA 15260}
\date{March 9, 1997}
\maketitle

\begin{abstract}
We study the effects of weak point and columnar disorder on the
vortex-lattice phase transitions in high temperature superconductors. The
combined effect of thermal fluctuations and of quenched disorder is
investigated using a simplified cage model. For point disorder we use the
mapping to a directed polymer in a disordered medium in 2+1 dimensions. For
columnar disorder the problem is mapped into a quantum particle in a
harmonic + random potential. We use the variational approximation to show
that point and columnar disorder have opposite effect on the position of the
melting line as is observed experimentally. For point disorder, replica
symmetry breaking plays a role at the transition into a vortex glass at low
temperatures.
\end{abstract}


\newpage

\section{Introduction}

There is a lot of interest in the physics of high temperature
superconductors due to their potential technological applications. In
particular these materials are of type II and allow for partial magnetic
flux penetration. Pinning of the magnetic flux lines (FL) by many types of
disorder is essential to eliminate dissipative losses associated with flux
motion. In clean materials below the superconducting temperature there exist
a 'solid ' phase where the vortex lines form a triangular Abrikosov lattice 
\cite{blatter}. In the presence of impurities it was suggested \cite{BG} the
Abrikosov crystal is replaced by a dislocation -free 'Bragg glass' which is
also characterized by (quasi-) long range order. This 'solid' can melt
due to thermal 
fluctuations or changes in the magnetic field. In particular known observed
transitions are into a flux liquid at higher temperatures via a first-order 
{\it melting line} (ML)\cite{zeldov}, and into a vortex glass at low
temperature \cite{VG},\cite{Fisher}, in the presence of disorder- the so
called {\it entanglement line} (EL). \cite{blatter}

Recently the effect of point and columnar disorder on the position of the
melting transition has been measured experimentally in the high-$T_c$
material $Bi_2Sr_2CaCu_2O_8$ \cite{Khaykovitch}(BSCCO). Point disorder has
been induced by electron irradiation (with 2.5 MeV electrons), whereas
columnar disorder has been induced by heavy ion irradiation (1 GeV Xe or 0.9
GeV Pb). It turns out that the flux melting transition persists in the
presence of either type of disorder, but its position shifts depending on
the disorder type and strength.

A significant difference has been observed between the effects of columnar
and point disorder on the location of the ML. Weak columnar defects
stabilize the solid phase with respect to the vortex liquid phase and shift
the transition to {\it higher} fields, whereas point-like disorder
destabilizes the vortex lattice and shifts the melting transition to {\it %
lower} fields. In this paper we attempt to provide a quantitative
explanation to this observation. The case of point defects has been
addressed in a recent paper by Ertas and Nelson \cite{EN} using the
cage-model approach which replaces the effect of vortex-vortex interactions
by an harmonic potential felt by a single vortex. For columnar disorder the
parabolic cage model was introduced by Nelson and Vinokur \cite{nelson}.
Here we use an analytic approach to analyze the cage-model Hamiltonian vis.
the replica method together with the variational approximation. In the case
of columnar defects our approach relies on our recent analysis of a quantum
particle in a random potential \cite{yygold,chen}. We compare the effect of the
two types of disorder with each other and with results of recent experiments.

Assume that the average magnetic field is aligned along the $z$-axis which
is also the c-axis of BSCCO, i.e. perpendicular to the CuO planes. Following
EN we describe the Hamiltonian of a single FL whose position is given by a
two-component vector ${\bf r}(z)$ (overhangs are neglected) by: 
\begin{eqnarray}
{\cal H}=\int_0^Ldz\left\{ {\frac{\epsilon _l}2}\left( {\frac{d{\bf r}}{dz}}%
\right) ^2+V(z,{\bf r})+{\frac \mu 2}{\bf r}^2\right\} .  \label{hamil}
\end{eqnarray}

Here $\epsilon _l=\epsilon _0/\gamma ^2$ is the line tension of the FL, $%
\gamma ^2=m_z/m_{\perp }$ is the mass anisotropy, $\epsilon _0=(\Phi _0/4\pi
\lambda )^2$, ( $\Phi _0$ is the fluxoid and $\lambda $ is the penetration
length), and $\mu \approx \epsilon _0/a_0^2$ is the effective spring
constant (setting the cage size) due to interactions with neighboring FLs,
which are at a typical distance of $a_0=\sqrt{\Phi _0/B}$ apart.

For the case of point-disorder, $V$ depends on $z$ and \cite{EN} 
\begin{eqnarray}
\left\langle V(z,{\bf r})V(z^{\prime },{\bf r^{\prime }})\right\rangle =%
\tilde \Delta \epsilon _0^2\xi ^3\delta _\xi ^{(2)}({\bf r}-{\bf r^{\prime }}%
)\delta (z-z^{\prime }).  \label{VVP}
\end{eqnarray}
where 
\begin{eqnarray}
\delta _\xi ^{(2)}({\bf r}-{\bf r^{\prime }})\approx 1/(2\pi \xi ^2)\exp (-(%
{\bf r}-{\bf r^{\prime }})^2/2\xi ^2),  \label{delta}
\end{eqnarray}
and $\xi $ is the vortex core diameter. The dimensionless parameter $\tilde 
\Delta $ is a measure of the strength of the disorder. In this case the
Hamiltonian is exactly the same as that of a directed polymer in a random
medium in 2+1 dimensions \cite{mp}. The vortex line represents the polymer
which is directed along the z-axis (overhangs are neglected). The presence
of impurities enhances the transverse wandering of the polymer.

For the case of columnar (or correlated) disorder, $V(z,{\bf r})=V({\bf r})$
is independent of $z$, and 
\begin{eqnarray}
\langle V({\bf r})V({\bf r^{\prime }})\rangle \equiv -2f(({\bf r}-{\bf %
r^{\prime }})^2/2)=g\epsilon _0^2\xi ^2\delta _\xi ^{(2)}({\bf r}-{\bf %
r^{\prime }}),  \label{VVC}
\end{eqnarray}
In this case we map the problem of a FL in a superconductor into that of a
quantum particle in a random potential. The partition function of the
quantum particle is just like the partition sum of the FL, provided one make
the identification \cite{nelson} 
\begin{eqnarray}
\hbar \rightarrow T,\qquad \beta \hbar \rightarrow L,  \label{corresp}
\end{eqnarray}
Where $T$ is the temperature of the superconductor and $L$ is the system
size in the $z$-direction. $\beta $ is the inverse temperature of the
quantum particle. We are interested in large fixed $L$ as $T$ is varied,
which corresponds to high $\beta $ for the quantum particle when $\hbar $
(or alternatively the mass of the particle) is varied. The variable $z$ is
the so called Trotter time.

The quantity which measures the transverse excursion of the FL is 
\begin{eqnarray}
u_0^2(\ell )\equiv \langle |{\bf r}(z)-{\bf r}(z+\ell )|^2\rangle \ /2,
\label{ul}
\end{eqnarray}
The main effect of the harmonic (or cage) potential is to cap the transverse
excursions of the FL beyond a confinement length $\ell ^{*}\approx
a_0/\gamma $. This length arises by equating the elastic energy and the cage
potential energy of the FL. Typically after it wanders a distance $\ell ^{*}$
along the z-direction the FL is reflected back by the walls of the cage and
restarts its transverse excursions. The near saturation of $u_0^2(\ell )$ at 
$\ell =\ell ^{*}$ will become evident from the analytical expressions
derived in the following sections. We thus define the mean square
displacement of the flux line by

\begin{equation}
u^2(T)=u_0^2(\ell ^{*}).  \label{uT}
\end{equation}

The location of the melting line is determined by the Lindemann criterion 
\begin{equation}
u^2(T_m(B))=c_L^2a_0^2,  \label{Lind}
\end{equation}
where $c_L\approx 0.15-0.2$ is the phenomenological Lindemann constant. This
means that when the transverse excursion of a section of length $\approx
\ell ^{*}$becomes comparable to a finite fraction of the interline
separation $a_0$, the melting of the flux solid occurs.

\section{The case of point disorder}

We start with the case of point disorder that is simpler mathematically. In
this case, the problem is equivalent to a directed polymer in a combination
of a random potential and a fixed harmonic potential. The flux line plays
the role of the polymer directed along the z-axis. The cage potential
supplies the harmonic part of the potential and the defects, or impurities,
the random part. The problem of directed polymers has been investigated
extensively in the literature in the absence of the harmonic piece. Here we
follow the approach of Mezard and Parisi (MP) \cite{mp}, who used the so
called variational (or Hartree) approximation. They set up the problem in
the presence of a harmonic piece with spring constant $\mu $, but they were
mainly concerned with the limit of $\mu \rightarrow 0$.

Recall that the Hamiltonian ${\cal H\,}$representing the system is given
(within the framework of the cage model) by eq. (\ref{hamil}), together with
probability distribution for the random potential whose second moment is
given by:

\begin{equation}
\left\langle V(z,{\bf r})V(z^{\prime },{\bf r^{\prime }})\right\rangle =%
\tilde \Delta \epsilon _0^2\xi ^3\delta _\xi ^{(2)}({\bf r}-{\bf r^{\prime }}%
)\delta (z-z^{\prime }).  \label{vvpd}
\end{equation}

In order to average over the quenched random potential, the replica method
is used. After introducing $n$ copies of the fields and averaging over the
potential one obtains:

\begin{equation}
\left\langle Z^n\right\rangle =\int d[{\bf r}_1]\dots d[{\bf r}_n]\
exp(-\beta H_n)\ ,
\end{equation}
with the replicated $n$-body Hamiltonian given by:

\begin{equation}
H_n={\frac{\epsilon _l}2}\int_0^Ldz\sum_{a=1}^n\left( \frac{d{\bf r}_a}{dz}%
\right) ^2+{\frac \mu 2}\int dz\ \sum_{a=1}^n\left( {\bf r}_a({z})\right) ^2-%
{\frac \beta 2}\frac{\tilde \Delta }{2\pi }\xi \epsilon _0^2\int d{z}\
\sum_{a,b}{\rm \exp }(-\frac{({\bf r}_a-{\bf r}_b)^2}{2\xi ^2})\ ,
\label{Hn}
\end{equation}
The variational quadratic Hamiltonian associated with the replica
Hamiltonian $H_n{\cal \,}$is parametrized by:

\begin{eqnarray}
h_n &=&{\frac 12}\int_0^Ldz\sum_a[\epsilon _l\ {\bf \dot r}_a^2+\mu {\bf r}%
_a^2]  \nonumber \\
&&\ \ \ -\frac 12\int_0^Ldz\sum_{a,b}^{}s_{ab}\ {\bf r}_a(z)\cdot {\bf r}%
_b(z),
\end{eqnarray}

where $s_{ab}\,$is the $n\times n\ $matrix of parameters needed to be
determined by the variational principles.

These parameters are fixed by extremizing the variational free-energy given
by: 
\begin{equation}
F=\left\langle H_n-h_n\right\rangle _{h_n}-\frac 1\beta \ln \left( \int d[%
{\bf r}_a]\ \exp (-\beta h_n)\right) .  \label{vfe}
\end{equation}

This free energy is given by: 
\begin{eqnarray}
\frac F{2L} &=&\frac 1{2\beta }\sum_{ab}s_{ab}\int_{-\infty }^\infty \frac{%
d\omega }{2\pi }G_{ab}(\omega )-\frac 1{2\beta }\int \frac{d\omega }{2\pi }%
{\rm Tr}\ln G(\omega )  \nonumber \\
&&-\frac \beta 4\frac{\tilde \Delta }{2\pi }\xi \epsilon _0^2\
\sum_{a,b}\sum_{m=0}^\infty \frac{(-)^m}{2^m\xi ^{2m}m!}\left\langle ({\bf r}%
_a(z)-{\bf r}_b(z))^{2m}\right\rangle _{h_n},  \label{vfe2}
\end{eqnarray}

where $G(\omega )\ $is the propagator associated with $h_{n\ }$: 
\begin{equation}
\lbrack G_{}^{-1}(\omega )]_{ab}=(\epsilon _l\ \omega ^2+\mu )\delta
_{ab}-s_{ab},  \label{pr}
\end{equation}

and $\omega $ is the 'momentum' variable conjugate to $z$. Since we are
interested in the limit $L\rightarrow \infty $ we can assume that $\omega $
is a continuous variable. Using the formula (for a two dimensional vector
field ${\bf r}$\ ): 
\begin{equation}
\left\langle ({\bf r}_a(z)-{\bf r}_b(z))^{2m}\right\rangle _{h_n}=\frac{2^mm!%
}{\beta ^m}\left( \int \frac{d\omega }{2\pi }[G_{aa}(\omega )+G_{bb}(\omega
)-2G_{ab}(\omega )]\right) ^m,
\end{equation}

we finally arrive at the result: 
\begin{eqnarray}
\frac F{2L} &=&const.+\frac 1{2\beta }\int \frac{d\omega }{2\pi }(\epsilon
_l\ \omega ^2+\mu )\sum_aG_{aa}(\omega )-\frac 1{2\beta }\int \frac{d\omega 
}{2\pi }{\rm Tr}\ln G(\omega )  \nonumber \\
&&\ +\frac \beta 2\sum_{ab}\widehat{f}_p(\frac 1\beta \int \frac{d\omega }{%
2\pi }[G_{aa}(\omega )+G_{bb}(\omega )-2G_{ab}(\omega )]),  \label{fvfe}
\end{eqnarray}

where the function $\widehat{f}_p\ $is given by:

\begin{equation}
\widehat{f}_p(y)=-\frac{\widetilde{\Delta }\epsilon _0^2\xi ^3}{4\pi }\frac 1%
{\xi ^2+y}  \label{fp}
\end{equation}

Stationarity of the free energy with respect to the parameters $s_{ab}$\
gives: 
\begin{eqnarray}
s_{ab} &=&2\beta \widehat{f}_p^{\prime }(\frac 1\beta \int \frac{d\omega }{%
2\pi }[G_{aa}(\omega )+G_{bb}(\omega )-2G_{ab}(\omega )]),\ \ \ a\neq b 
\nonumber \\
s_{aa} &=&-\sum_{b(\neq a)}s_{ab.}  \label{sc}
\end{eqnarray}
Here $\widehat{f}_p^{\prime }$ is the derivative of $\widehat{f}_p(y)$ with
respect to its argument.

We consider first the replica symmetric (RS) case, where all the off
diagonal elements of $s_{ab}$ are taken to be equal to each other and their
value denoted by $s$. We denote the value of the diagonal elements by $s_d$.
Eq.(\ref{sc}) implies that in the limit $n\rightarrow 0$, $s_d=s$.

In this limit we find:

\begin{eqnarray}
G_{ab}(\omega ) &=&\frac{\delta _{ab}}{\epsilon _l\ \omega ^2+\mu }+\frac s{%
(\epsilon _l\ \omega ^2+\mu )^2}, \\
s &=&\frac 2T\widehat{f}_p\ ^{\prime }(\tau ).
\end{eqnarray}
Here we introduced the reduced temperature variable 
\begin{equation}
\tau =T/\sqrt{\epsilon _l\mu }.  \label{tau}
\end{equation}
Using these results we can calculate the mean square displacement (\ref{ul}%
): 
\begin{eqnarray}
u_0^2(\ell ) &=&\frac 2\beta \int \frac{d\omega }{2\pi }(1-\cos (\omega \ell
))G_{aa}(\omega )  \nonumber \\
&=&2T\int \frac{d\omega }{2\pi }\frac{1-\cos (\omega \ell )}{\epsilon _l\
\omega ^2+\mu }(1+\frac{s^{}}{\epsilon _l\ \omega ^2+\mu }),
\end{eqnarray}
and hence

\begin{eqnarray}
u_0^2(\ell ) &=&\tau (1-e^{-\ell /\ell ^{*}})+\tau \ s\ /\ (2\mu )  \nonumber
\\
&&\ \ \ \ \ \times \ (1-e^{-\ell /\ell ^{*}}-(\ell /\ell ^{*})\ e^{-\ell
/\ell ^{*}}),
\end{eqnarray}
with 
\begin{equation}
s=\frac{\epsilon _0^2\xi ^3}T\frac{\widetilde{\Delta }}{2\pi }\frac 1{(\xi
^2+\tau )^2},\ \ \ell ^{*}=\sqrt{\epsilon _l/\mu }.
\end{equation}

Note that $s$ is positive and independent of $\omega $, and hence the mean
square displacement $u_0^2(\ell ^{*})$ is bigger than its value for zero
disorder.

At this point it is convenient to introduce dimensionless variables: 
\begin{eqnarray}
\widetilde{T} &=&T/(\epsilon _0\xi ),  \label{Tda} \\
\widetilde{B} &=&B\xi ^2/\Phi _0.  \label{Bda}
\end{eqnarray}
In terms of these variables 
\begin{eqnarray}
\mu &=&\widetilde{B}\ \epsilon _0/\xi ^2,\ \ \ a_0^2=\xi ^2/\widetilde{B}, \\
\tau &=&\xi ^2\gamma \widetilde{T}\ /\ \sqrt{\widetilde{B}}.
\end{eqnarray}
As parameters for BSCCO we take mean values for those quoted in ref. \cite
{blatter}: 
\begin{eqnarray}
\xi &\cong &30\AA ,  \nonumber \\
\epsilon _0\xi &\cong &1905K,\ \text{(corresp. to }\lambda _L\approx 1700\AA 
\text{)}  \label{BSCCO} \\
\gamma &\cong &125.  \nonumber
\end{eqnarray}
and also $\Phi _0=2.07\times 10^{-7}$ G cm$^2.$ In the case of the
experiments one has

\begin{equation}
\widetilde{B}\ll (\gamma \widetilde{T})^2,  \label{BvsT}
\end{equation}
and hence: 
\begin{equation}
u_{}^2(T)\ /\ a_0^2\simeq \sqrt{\widetilde{B}}\gamma \widetilde{T}\ (\
1-e^{-1})+\frac 12\sqrt{\widetilde{B}}%
{\gamma \widetilde{\Delta } \overwithdelims() 2\pi }
\frac 1{(\gamma \widetilde{T})^2}\ (1-2e^{-1})  \label{uTrs}
\end{equation}
Fig. 1 curve {\it a }shows a plot of $\sqrt{u_0^2(\ell ^{*})}/a_0$ vs. $T$\
for $\widetilde{\Delta }/2\pi =0$ (curve a) and for $\widetilde{\Delta }%
/2\pi =0.2$ (curve b). The value of the magnetic field is taken to be $%
B=250G.$

Using the Lindemann criterion, eq. (\ref{Lind}), we can easily solve for the
magnetic field at the melting transition: 
\begin{equation}
\widetilde{B}_m(\widetilde{T})=c_L^4\times \left( \gamma \widetilde{T}\ (\
1-e^{-1})+\frac 12%
{\gamma \widetilde{\Delta } \overwithdelims() 2\pi }
\frac 1{(\gamma \widetilde{T})^2}\ (1-2e^{-1})\right) ^{-2}  \label{Bm}
\end{equation}

For $T<T_{cp}$ with 
\begin{equation}
T_{cp}\approx (\epsilon _0\xi /\gamma )(\gamma \widetilde{\Delta }/2\pi
)^{1/3}  \label{Tcp}
\end{equation}
it is necessary to break replica symmetry. This means that the off-diagonal
elements of the variational matrix $s_{ab}$ are not all equal to each other.
MP \cite{mp} worked out the equations for the replica symmetry breaking
(RSB) solution in the limit of $\mu \rightarrow 0$, but it is not difficult
to extend them to any value of $\mu .$ Their final solution is not
applicable for the particular correlation $\widetilde{f}_p$ discussed in
this paper, hence we will work it out below and in the Appendix.

When breaking replica symmetry {\it a la }Parisi, it is customary to
introduce Parisi's parameter which is denoted here by $u$. Thus we put 
\begin{eqnarray}
s_{aa} &=&s_d, \\
a &\neq &b,\ \ s_{ab}=s(u),\ \ 0\leq u\leq 1.
\end{eqnarray}
We have found that a 1-step RSB solution is sufficient in the present case
and we parametrize it by: 
\begin{eqnarray}
s(u) &=&\left\{ 
{s_0\ \ \ \ u<u_c \atop s_1\ \ \ \ u>u_c}
\right.  \label{su} \\
\Sigma &=&u_c(s_1-s_0).  \label{Sig}
\end{eqnarray}
In the following we will used the dimensionless variable $\widetilde{s}=s\xi
^2/\epsilon _0$, and similarly for $\widetilde{\Sigma }$. In the case of the
experiments, the assumption of small $\mu $ amounts to the condition (\ref
{BvsT}) which is very well satisfied. For small $\mu $ we find (see Appendix
for further details): 
\begin{eqnarray}
u_c &=&\widetilde{T}/\widetilde{T}_{cp},  \label{ucs} \\
\widetilde{\Sigma } &=&(\gamma \widetilde{T}_{cp}-\gamma \widetilde{T})^2,
\label{Sigs} \\
s_0 &=&\frac 1{\gamma \widetilde{T}}\frac{\widetilde{B}}{(\gamma \widetilde{T%
}_{cp})^2}%
{\gamma \widetilde{\Delta } \overwithdelims() 2\pi }
,  \label{s0s} \\
\gamma \widetilde{T}_{cp} &=&%
{\gamma \widetilde{\Delta } \overwithdelims() 2\pi }
^{1/3}.  \label{Tcps}
\end{eqnarray}
Below $T_{cp}$ the mean square displacement freezes at its value at $T_{cp}:$
\begin{equation}
u_0^2(T)\simeq u_0^2(T_{cp}),\ \ \ T\leq T_{cp}  \label{u2rsb}
\end{equation}
The 'freezing' temperature $T_{cp\text{ }}$is about 1.33 larger than the
temperature for which the RS expression for $u_0^2(T)$ (see equation (\ref
{uTrs}) has a minimum. The value of the magnetic field corresponding to $%
T_{cp\text{ }}$is $B_m(T_{cp})\approx 2.07(\Phi _0/\xi ^2)(\gamma \widetilde{%
\Delta }/2\pi )^{-2/3}c_L^4$ gives a reasonable agreement with the
experiments as is evident from Figure 2.

In Figure 2 we show a plot of $B_m(T)$ vs. $T$ for different values of the
disorder. The points represent data reported by B. Khaykovitch {\it et al. }%
\cite{Khaykovitch} for various amounts of point disorder (we also display
points for columnar disorder as will be discussed in the next section). The
theoretical fit is done using equation (\ref{Bm}) for $T>T_{cp}$ and $%
B_m(T)=B_m(T_{cp})$, for $T<T_{cp}$. Best fit has been obtain by choosing $%
c_L=0.162$ and $\widetilde{\Delta }/2\pi =$ 0.144, 0.208 and 0.280
respectively. We have chosen the amount of disorder to best fit the
temperature $T_{cp}$ below which the experimental curves show an apparent
change of behavior. This is achieved by using eq. (\ref{Tcp}). The fit
associated with the value of $\widetilde{\Delta }/2\pi =$ 0.144 is the one
corresponding to the 'as grown' crystal which always has some amount of
point defects. At high temperatures one observe somewhat larger deviations
between the theoretical fits and the experimental data. This is due to the
proximity to $T_c\approx 90K$, which affects the behavior of the melting
line even for pure samples, see discussion in the concluding section. The
flat part of the curves represent the so called {\it entanglement line},
which is believed to be a (continuous) transition into the vortex glass.

\section{The case of columnar disorder}

We consider first the case of columnar disorder. This problem maps into the
problem of a quantum particle in a random potential. to see this we recall
that the density matrix at finite temperature (= $\beta ^{-1}$) of a quantum
particle subject to a potential V is given by: 
\begin{eqnarray}
\rho ({\bf r,r^{\prime }},U)\ =\ \int_{{\bf r}(0)={\bf r}}^{{\bf r}(U)={\bf r%
}^{\prime }}[d{\bf r}]\ \exp \left\{ -{\frac 1\hbar }\int_0^U\left[ {\frac{m%
\dot {{\bf r}}(z)^2}2}\ +\ {\frac{\mu {\bf r}(z)^2}2}\ +\ V({\bf r}%
(z))\right] dz\right\} ,  \label{densitym}
\end{eqnarray}
with $U=\beta \hbar $.The variable $z$ has dimensions of time and is often
referred to as the Trotter dimension. This is the same as the partition
function of a single flux line, provided one makes the identification 
\begin{eqnarray}
\hbar \rightarrow T,\ \ U=\beta \hbar \rightarrow L,\ \ m\rightarrow
\epsilon _l  \label{corresp2}
\end{eqnarray}
mentioned in the Introduction.

In the absence of disorder it is easily obtained from standard quantum
mechanics and the correspondence (\ref{corresp}), that when $L\rightarrow
\infty ,$

\begin{equation}
u^2(T)=\frac T{\sqrt{\epsilon _l\mu }}\left( 1-\exp (-\ell ^{*}\sqrt{\mu
/\epsilon _l})\right) =\frac T{\sqrt{\epsilon _l\mu }}(1-e^{-1}),
\label{u2g0}
\end{equation}

When we turn on disorder we have to solve the problem of a quantum particle
in a random quenched potential. This problem has been recently solved using
the replica method and the variational approximation \cite{yygold}. Let us
review briefly the results of this approach. We apply the replica trick in
order to carry out the quenched average over the random realizations. We
consider $n$-copies of the system, and obtain for the averaged density
matrix: 
\begin{eqnarray}
\rho ({\bf r}_1\cdots {\bf r}_n,{\bf r}_1\cdots {\bf r}_n,L)\ &=&\ \int_{%
{\bf r}_a(0)={\bf r}_a}^{{\bf r}_a(L)={\bf r}_a}\ \prod_{a=1}^n[d{\bf r}%
_a]\exp \left\{ -{\cal H}_n/T\right\} , \\
{\cal H}_n\ &=&\ {\frac 12}\int_0^Ldz\ \sum_a\left[ \epsilon _l{\bf r}%
_a^2(u)+\ \mu {\bf r}_a^2(u)\right] \hspace{1.75in}  \nonumber \\
&&\ \ +\ {\frac 1{2T}}\int_0^Ldz\int_0^Ldz^{\prime }\sum_{ab}2\ f\ \left( {%
\frac{({\bf r}_a(z)-{\bf r}_b(z^{\prime }))^2}2}\right) ,  \label{Hnc}
\end{eqnarray}
with 
\begin{equation}
f(y)=-\frac{g\epsilon _0^2}{4\pi }\exp (-\frac y{\xi ^2})
\end{equation}

In this approximation we chose the best quadratic Hamiltonian parametrized
by the matrix $s_{ab}(z-z^{\prime })$:

\begin{eqnarray}
h_n &=&\frac 12\int_0^Ldz\sum_a[\epsilon _l{\bf \dot r}_a^2+\mu {\bf r}_a^2]
\nonumber \\
&&\ -\frac 1{2T}\int_0^Ldz\int_0^Ldz^{\prime }\sum_{a,b}s_{ab}(z-z^{\prime })%
{\bf r}_a(z)\cdot {\bf r}_b(z^{\prime }).  \label{hn}
\end{eqnarray}
Here the replica index $a=1\ldots n$, and $n\rightarrow 0$ at the end of the
calculation. Again, this Hamiltonian is determined by stationarity of the
variational free energy which is given by

\begin{equation}
\left\langle F\right\rangle _R/T=\left\langle H_n-h_n\right\rangle
_{h_n}-\ln \int [d{\bf r}]\exp (-h_n/T),  \label{FV}
\end{equation}
The off-diagonal elements of $s_{ab}$ can consistently be taken to be
independent of $z$, whereas the diagonal elements are $z$-dependent. It is
more convenient to work in frequency space, where $\omega $ is the frequency
conjugate to $z$. $\omega _j=(2\pi /L)j$, with  $j=0,\pm 1,\pm 2,\ldots $.
Assuming replica symmetry, which is valid only for part of the temperature
range, we can denote the off-diagonal elements of $\widetilde{s}_{ab}(\omega
)=(1/T)\int_0^Ldz\ e^{i\omega z}$ $s_{ab}(z)$, by $\widetilde{s}(\omega )=%
\widetilde{s}\delta _{\omega ,0}$. Denoting the diagonal elements by $%
\widetilde{s}_d(\omega )$, the variational equations become: 
\begin{eqnarray}
\tilde s &=&2\frac LT\widehat{f_c}\ ^{\prime }\left( {\frac{2T}{\mu L}}+{%
\frac{2T}L}\sum_{\omega ^{\prime }\neq 0}\frac 1{\epsilon _l\ \omega
^{\prime }\,^2+\mu -\widetilde{s}_d(\omega ^{\prime })}\right)   \label{s} \\
\tilde s_d(\omega ) &=&\tilde s-{\frac 2T}\int_0^Ld\zeta \ (1-e^{i\omega
\zeta })\times   \nonumber \\
&&\ \ \ \widehat{f_c}\ ^{\prime }\left( {\frac{2T}L}\sum_{\omega ^{\prime
}\neq 0}\ \frac{1-e^{-i\omega ^{\prime }\varsigma }}{\epsilon _l\omega
^{\prime }\,^2+\mu -\widetilde{s}_d(\omega ^{\prime })}^{}\right) .
\label{sd}
\end{eqnarray}
here $\widehat{f_c}$ $^{\prime }(y)$ denotes the derivative of the
''dressed'' function $\widehat{f_c}(y)$ which is obtained in the variational
scheme from the random potential's correlation function $f(y)$, and in our
case is given by:

\begin{equation}
\widehat{f_c}(y)=-\frac{g\epsilon _0^2\xi ^2}{4\pi }\frac 1{\xi ^2+y}
\label{fc}
\end{equation}
The full equations, taking into account the possibility of replica-symmetry
breaking are given in ref. \cite{yygold}. In terms of the variational
parameters the function $u_0^2(\ell )$ is given by

\begin{equation}
u_0^2(\ell )={\frac{2T}L}\sum_{\omega ^{\prime }\neq 0}\frac{1-\cos (\omega
^{\prime }\ell ^{*})}{\epsilon _l\omega ^{\prime }\,^2+\mu -\widetilde{s}%
_d(\omega ^{\prime })}.  \label{u2qp}
\end{equation}
This quantity has not been calculated in ref. \cite{yygold}. There we
calculated $\left\langle {\bf r}^2(0)\right\rangle $ which does not measure
correlations along the $z$-direction.

In the limit $L\rightarrow \infty $ we were able to solve the equations
analytically to leading order in $g$. In that limit eq. (\ref{sd}) becomes
(for $\omega \neq 0$) :

\begin{eqnarray}
\tilde s_d(\omega ) &=&\frac 4\mu \widehat{f}_c\ ^{\prime \prime }(b_0)-%
\frac 2T\int_0^\infty d\varsigma (1-\cos (\omega \varsigma ))  \nonumber \\
&&\times (\widehat{f}_c\ ^{\prime }(C_0(\varsigma ))-\widehat{f}_c\ ^{\prime
}(b_0)),  \label{sdi}
\end{eqnarray}
with

\begin{eqnarray}
C_0(\varsigma ) &=&2T\int_{-\infty }^\infty \frac{d\omega }{2\pi }\frac{%
1-\cos (\omega \varsigma )}{\epsilon _l\omega \,^2+\mu -\widetilde{s}%
_d(\omega )},  \label{C0} \\
b_0 &=&2T\int_{-\infty }^\infty \frac{d\omega }{2\pi }\frac 1{\epsilon
_l\omega \,^2+\mu -\widetilde{s}_d(\omega )}.  \label{b0}
\end{eqnarray}

One can solve equation (\ref{sdi}) numerically, but there it is hard to
obtain good accuracy at high frequencies when the cosine term oscillates
strongly. We can get a better approximation analytically. We parametrize an
approximate solution by: 
\begin{equation}
\widetilde{s}_d(\omega )=s_\infty +A\mu /(\epsilon _l\omega ^2+a^2\mu ),\ \
\ (\omega \neq 0)  \label{sdp}
\end{equation}
and require that it will obey the correct behavior at low and high
frequencies to leading order in the strength of the disorder. There are
three parameters which are determined by $\widetilde{s}_d(\omega =0),\ 
\widetilde{s}_d^{\prime }(\omega =0)$, and $\widetilde{s}_d(\omega =\infty )$%
. To leading order in the strength of the disorder we can substitute in eq. (%
\ref{sdi}), 
\begin{equation}
b_0=\tau ,\ C_0(\zeta )=\tau (1-\exp (-|\zeta |\sqrt{\mu /\epsilon _l}\ ),
\end{equation}
with $\tau =T\ /\sqrt{\epsilon _l\ \mu }$. We then find after some algebra 
\begin{eqnarray}
s_\infty &=&\frac 4\mu \widehat{f}_c\ ^{\prime \prime }(\tau )(1+\frac 14%
\int_0^\infty d\zeta \ e^{-\zeta }(\frac 1{1-\alpha e^{-\zeta }}+\frac 1{%
(1-\alpha e^{-\zeta })^2}))  \nonumber \\
\ &=&\frac 1\mu \widehat{f}_c\ ^{\prime \prime }(\tau )(4+f_1(\alpha )),
\end{eqnarray}
where we defined

\begin{eqnarray}
\ \alpha  &=&\tau \ /(\xi ^2+\tau ),  \label{tau,al} \\
f_1(\alpha ) &=&1/(1-\alpha )-(1/\alpha )\log (1-\alpha ).  \label{f1}
\end{eqnarray}
Similarly for small $\omega $ we find: 
\begin{eqnarray}
s_d(\omega ) &=&\frac 4\mu \widehat{f}_c\ ^{\prime \prime }(\tau )(1+\omega
^2\frac 18\frac{\epsilon _l^{}}\mu \int_0^\infty d\zeta \ \zeta ^2e^{-\zeta
}(\frac 1{1-\alpha e^{-\zeta }}+\frac 1{(1-\alpha e^{-\zeta })^2})+\ldots ) 
\nonumber \\
\  &=&\frac 4\mu \widehat{f}_c\ ^{\prime \prime }(\tau )(1+\omega ^2\frac 14%
\frac{\epsilon _l^{}}\mu f_2(\alpha )+\ldots ),
\end{eqnarray}
with 
\begin{equation}
f_2(\alpha )=\frac 1\alpha \sum_{k=1}^\infty \frac{k+1}{k^3}\alpha ^k.
\end{equation}
From this equation we find for the other two parameters in eq. (\ref{sdp}) 
\begin{eqnarray}
a^2 &=&f_1(\alpha )/f_2(\alpha ), \\
A &=&-\frac{\widehat{f}_c\ ^{\prime \prime }(\tau )}\mu \frac{f_1^2(\alpha )%
}{f_2(\alpha )}.
\end{eqnarray}
Notice that $s_d(\omega )$ is negative for all $\omega >0$. It interpolates
from the value $-4|\ \widehat{f}_c\ ^{\prime \prime }(\tau )|$ /$\mu $ at $%
\omega \sim 0$ to the value $-(4+f_1(\alpha ))|\ \widehat{f}_c\ ^{\prime
\prime }(\tau )|$ /$\mu $ at $\omega =\infty $. Substituting (\ref{sdp}) in
eq. (\ref{C0}) and expanding the denominator to leading order in the
strength of the disorder, we get :

\begin{eqnarray}
u_0^2(\ell ) &=&C_0(\ell )=\tau (1-A/(a^2-1)^2/\mu )  \nonumber \\
&&\ \ \times (1-e^{-\ell /\ell ^{*}})+\tau A/(a(a^2-1)^2\mu )\times 
\nonumber \\
&&\ (1-e^{-a\ell /\ell ^{*}})+\tau /(2\mu )\times \ (s_\infty +A/(a^2-1)) 
\nonumber \\
&&\ \times \ (1-e^{-\ell /\ell ^{*}}-(\ell /\ell ^{*})\ e^{-\ell /\ell
^{*}}).
\end{eqnarray}
Recall that this result is valid to first order in the strength of the
columnar disorder.

This expression simplifies significantly under the assumption 
\begin{equation}
\widetilde{B}\ll (\gamma \widetilde{T})^2,
\end{equation}

valid in the experiments. In that case 
\begin{eqnarray}
\alpha  &\simeq &1-\sqrt{\widetilde{B}\ }/\ \gamma \widetilde{T}+\cdots , \\
a^2 &\simeq &0.351\gamma \widetilde{T}\ /\ \sqrt{\widetilde{B}}-0.475\ \log (%
\sqrt{\widetilde{B}\ }/\ \gamma \widetilde{T})+0.326+\cdots , \\
A/\mu  &\simeq &0.351\frac g{2\pi }\frac 1{\widetilde{B}^{3/2}(\gamma 
\widetilde{T}\ )}(1-(1.07+2.351\log (\sqrt{\widetilde{B}\ }/\ \gamma 
\widetilde{T}))\sqrt{\widetilde{B}\ }/\ \gamma \widetilde{T})+\cdots , \\
s_\infty /\mu  &=&-\frac g{2\pi }\frac 1{\widetilde{B}\ (\gamma \widetilde{T}%
\ )^2}(1+(2-\log (\sqrt{\widetilde{B}\ }/\ \gamma \widetilde{T}))\sqrt{%
\widetilde{B}\ }/\ \gamma \widetilde{T})+\cdots ,
\end{eqnarray}
from which we find: 
\begin{equation}
u^2(T)\ /\ a_0^2\simeq 0.632\sqrt{\widetilde{B}}\gamma \widetilde{T}\ -1.952%
{g \overwithdelims() 2\pi }
\frac 1{(\gamma \widetilde{T})^2}+4.804%
{g \overwithdelims() 2\pi }
\frac{\widetilde{B}^{1/4}}{(\gamma \widetilde{T}\ )^{5/2}}+\cdots \ .
\label{u2ac}
\end{equation}
From this equation we can derive the most important result for the location
of the melting transition: 
\begin{equation}
\widetilde{B}_m(\widetilde{T})=\frac{2.504c_L^4}{(\gamma \widetilde{T})^2}%
\left( 1+3.904c_L^{-2}%
{g \overwithdelims() 2\pi }
\frac 1{(\gamma \widetilde{T})^2}(1-\frac{3.093c_L}{\gamma \widetilde{T}}%
)\right) ^{}.  \label{Bmc}
\end{equation}
In Fig. 1 shows a plot of $\sqrt{u^2(T)}/a_0$ vs. $T$ for $g/2\pi =0.025$
(curve d). We have chosen $B=250G$. We see that the disorder tends to align
the flux lines along the columnar defects, hence decreasing $u^2(T)$.
Technically this happens since $\widetilde{s}_d(\omega )$ is negative.

For the case of combined columnar and point disorder, it is tempting to
combine equations (\ref{Bm}) and (\ref{Bmc}) into a single equation for $%
T>T_{cp}$: 
\begin{equation}
\widetilde{B}_m(\widetilde{T})=\frac{2.504c_L^4}{(\gamma \widetilde{T})^2}\ 
\frac{1+3.904c_L^{-2}%
{g \overwithdelims() 2\pi }
\frac 1{(\gamma \widetilde{T})^2}(1-\frac{3.093c_L}{\gamma \widetilde{T}})}{%
\left( 1+0.209%
{\gamma \widetilde{\Delta } \overwithdelims() 2\pi }
\frac 1{(\gamma \widetilde{T})^3}\right) ^2}^{}.  \label{Bmf}
\end{equation}
Even the naturally grown crystals has some amount of point disorder as
discussed above.

In Fig. 2 we show the modified melting line $B_m(T)$ in the presence of
columnar disorder and a small amount of point disorder ($\ \widetilde{\Delta 
}/2\pi =0.144$, corresponding to the 'as grown' curve fit of the last
section), as given by eq. (\ref{Bmf}) with $c_L=0.162$. We see that the
melting line shifts towards higher magnetic fields with increasing amounts
of columnar disorder. The best fit to the experimental results is obtained
for $g/2\pi =0.01$ and $g/2\pi =0.025$ respectively.

For $T<T_c\approx (\epsilon _0\xi /\gamma )[4g^2\Phi _0\ /\ (\pi ^2\xi
^2B)]^{1/6}$, there is a solution with RSB. This temperature is below the
bottom of the range plotted in the figures for columnar disorder. It is thus
not necessary to include the RSB solution in the plot. 

\section{Conclusions}

The analytical expressions given in eqs. (\ref{Bm}), (\ref{Bmc}), though
quite simple, seem to capture the essential feature required to reproduce
the position of the melting line. The qualitative agreement with
experimental results is remarkable, especially the opposite effects of point
and columnar disorder on the position of the melting line. The 'as grown'
experimental results are, as expected, corresponding to small amount of
point disorder.

The effect of point disorder is to increase the transverse excursions of the
FL which seeking the best free-energetic configuration. This increase in the
mean square fluctuations lowers the melting temperature for a given magnetic
field. At low temperature, the entanglement transition is associated in our
formalism with RSB, and is a sort of a spin-glass transition in the sense
that many minima of the random potential and hence free energy, compete with
each other. This means that there are many possible deformations of the FL
which are very close in energy, or free energy. To our knowledge this is the
first time the transition into the vortex glass is represented as a RSB
transition.

For point disorder, in the limit of infinite cage ( $\mu \rightarrow 0$),
the variational approximation gives a wandering exponent of 1/2 for a random
potential with short ranged correlations \cite{mp}, whereas simulations give
a value of 5/8 \cite{halpin}. This discrepancy does not seem of importance
with respect to the conclusions obtained in this paper since we always
consider the case of finite $\mu $ (which amount to a non-zero magnetic
field) and also consider distances of order $\ell ^{*}$ in the z-direction.

For columnar disorder, the z-independence of the random potential tends to
reduce the transverse fluctuations of the FL, thus shifting the melting
transition to higher temperatures and magnetic fields. Related to this point
is the fact that columnar disorder is much more effective in shifting the
position of the melting line as compared with point disorder for the range
of parameters considered here. We have used a much weaker value of
correlated disorder to achieve a similar or even larger shift of the melting
line than for the case of point disorder. This is again related to the
z-independence of the random potential in the columnar case, which help to
enhance its effect on the excursions of the vortex lines.

The experiments show that in the case of columnar disorder the transition
into the vortex glass seems to be absent. In our model it corresponds to the
fact that RSB does not occur in the temperature range of relevance to the
experiments, but rather at significantly lower temperatures for the amount
of disorder present.

Concerning the apparent deviations of the theoretical curves from the
experimental points at high temperatures, we should point out that our
expressions are only valid far from $Tc$ which is $\simeq $90K for BSCCO.
Close to $T_c$ the melting line cannot behave as $B_m\propto 1/T^2$ since it
must terminate at $B$=0, $T=T_c$. This comes about because of the
temperature dependence of the fundamental constants like the penetration and
coherence lengths. Houghton {\it et al.} \cite{houghton} used a detailed
melting theory to obtain a behavior of $B_m\approx B_0(1-T/T_c)^2$ near $Tc$%
. Other authors \cite{glazman}predict a behavior like $B_m\approx
B_0(T_c/T-1)$, which seems to better fit the experimental data for the
system under consideration according to ref. \cite{Khaykovitch}. Adding such
a correction factor to our expression for $B_m(T)$ can improve the fit at
high temperatures.

Another point we should mention is that our model incorporates the disorder
as depending on a single parameter like $\widetilde{\Delta }$ or $g$ ,
whereas experimentally there are two parameters which are the density of
defects, which is varied experimentally, and their individual strength over
which there is not much control. Our model is valid in the limit of very
weak impurities which are densely and uniformly distributed in space. This
is quite reasonable for the case of point disorder, but for columnar
disorder the experiments involve a rather low density of columnar defects
which is smaller than the density of vortices. Thus from the point of view
of a single vortex the disorder is not uniformly distributed which may
account for the apparent difference in curvature between theory and
experiment.

We have shown that the {\it cage model }together with the variational
approximation reproduce the main feature of the experiments. Effects of many
body interaction between vortex lines which are not taken into account by
the effective cage model seem to be of secondary importance. Inclusion of
such collective effects within the variational formalism remains a task for
the future. These effects may be responsible for the apparent difference in
curvature between the experimental and theoretical curves, for the case of
columnar disorder.

We thank David Nelson and Eli Zeldov for discussions. We thank the Weizmann
Institute for a Michael Visiting Professorship, during which this research
has been initiated.

\section{Appendix}

In this appendix we give more details of the RSB solutions for point
disorder and for columnar disorder. We use some of the results found in
Appendix II of MP (\cite{mp}).Using the parametrization given in eqs.(\ref
{su}) and (\ref{Sig}), we find for the free energy: 
\begin{eqnarray}
\frac F{2L} &=&const.+\frac \tau 4\frac{1-u_c^{}}{u_c}\Sigma \sqrt{\frac \mu
{\mu +\Sigma }}-\frac \tau 2\frac{1-u_c^{}}{u_c}\sqrt{\mu (\mu +\Sigma )} 
\nonumber \\
&&+\beta \frac \Delta {8\pi }u_c\frac 1{\xi ^2+\frac \tau {u_c}-\frac{%
1-u_c^{}}{u_c}\tau \sqrt{\frac \mu {\mu +\Sigma }}}+\beta \frac \Delta {8\pi 
}(1-u_c)\frac 1{\xi ^2+\tau \sqrt{\frac \mu {\mu +\Sigma }}},  \label{FRSB}
\end{eqnarray}
where we defined $\Delta =\widetilde{\Delta }\epsilon _0^2\xi ^3$.
Stationarity with respect to $\Sigma $ and $u_c$ yields equations for these
two quantities. Introducing the dimensionless quantity 
\begin{equation}
\widetilde{\Sigma }=\Sigma \xi ^2/\epsilon _0,
\end{equation}
and taking the limit of small $\mu ,$ these equations become: 
\begin{eqnarray}
\widetilde{\Sigma } &=&\frac{u_c}{\gamma \widetilde{T}}\frac{\gamma 
\widetilde{\Delta }}{2\pi }\frac 1{(1+\gamma \widetilde{T\ }/\sqrt{%
\widetilde{\Sigma }})^2}, \\
\ \sqrt{\widetilde{\Sigma }} &=&\frac{u_c^2}{(\gamma \widetilde{T})^2}\frac{%
\gamma \widetilde{\Delta }}{2\pi }\frac 1{(1+\gamma \widetilde{T\ }/\sqrt{%
\widetilde{\Sigma }})}
\end{eqnarray}
Solving these equations we find the solutions given in equations (\ref{ucs})
and (\ref{Sigs}). The quantities $s_0$ and $s_1$ are given by the equations: 
\begin{eqnarray}
s_0 &=&2\beta \frac \Delta {4\pi }\frac 1{(\xi ^2+\frac \tau {u_c}-\frac{%
1-u_c^{}}{u_c}\tau \sqrt{\frac \mu {\mu +\Sigma }})^2},  \label{s0a} \\
s_1 &=&2\beta \frac \Delta {4\pi }\frac 1{(\xi ^2+\tau \sqrt{\frac \mu {\mu
+\Sigma }})^2},  \label{s1a}
\end{eqnarray}
from which equation (\ref{s0s}) follows in the limit of small $\mu $.

The mean square displacement $u_0^2(\ell )$ is given by 
\begin{eqnarray}
u_0^2(\ell ) &=&2T\int \frac{d\omega }{2\pi }(1-\cos (\omega \ell
))G_{aa}(\omega )  \nonumber \\
\ &=&\frac \tau {u_c}(1-\exp (-\ell \sqrt{\mu /\epsilon _l}))-\frac{1-u_c^{}%
}{u_c}\tau \sqrt{\frac \mu {\mu +\Sigma }}(1-\exp (-\ell \sqrt{(\mu +\Sigma
)/\epsilon _l}))  \nonumber \\
&&\ +s_0\frac \tau {2\mu }(1-\exp (-\ell \sqrt{\mu /\epsilon _l})-(\ell 
\sqrt{\mu /\epsilon _l})\exp (-\ell \sqrt{\mu /\epsilon _l})).
\end{eqnarray}
From this expression we obtain eq.(\ref{u2rsb}) for small $\mu $.

Let us discuss briefly the case of columnar disorder. In this case we showed
in ref. \cite{yygold}that replica symmetry is broken in a region of the $%
T-\hbar ^2/m$ phase diagram. In the present paper we use the mapping given
by equation (\ref{corresp2}), and in addition we take the limit $%
L\rightarrow \infty $ . For the case of short ranged correlation of the
potential a 1-step RSB has been found. In the limit $L\rightarrow \infty $
we find that the breaking point $u_{c\text{ }}$of the one step solution
scales as $L^{-1}$. Nevertheless there is a finite contribution to
observable quantities as will be discussed below. Putting 
\begin{eqnarray}
\widetilde{s}(z) &=&\left\{ 
{\widetilde{s}_0\ \ \ \ u<u_c \atop \widetilde{s}_1\ \ \ \ u>u_c}
\right.  \label{suc} \\
\Sigma &=&u_c(\widetilde{s}_1-\widetilde{s}_0),  \label{Sigc} \\
u_c &=&y_c/L,
\end{eqnarray}
we find the following equation for the limit of large $L$: 
\begin{eqnarray}
\widetilde{s}_0/L &=&\frac 2T\widehat{f}_c\ ^{\prime }\ (b_0+\frac{2T\Sigma 
}{y_c\mu (\mu +\Sigma )}) \\
\widetilde{s}_1/L &=&\frac 2T\widehat{f}_c\ ^{\prime }\ (b_0) \\
\widetilde{s}_d(\omega ) &=&-\Sigma +\frac 4{\mu +\Sigma }\widehat{f}_c\
^{\prime \prime }\ (b_0)-\frac 2T\int_0^\infty d\varsigma (1-\cos (\omega
\varsigma ))  \nonumber \\
&&\ \times (\widehat{f}_c\ ^{\prime }(C_0(\varsigma ))-\widehat{f}_c\
^{\prime }(b_0)),
\end{eqnarray}
and $\Sigma $ and $y_c$ satisfying the equations 
\begin{eqnarray}
\Sigma &=&\frac{2y_c}T(\widehat{f}_c\ ^{\prime }\ (b_0)-\widehat{f}_c\
^{\prime }\ (b_0+\frac{2T\Sigma }{y_c\mu (\mu +\Sigma )}), \\
0 &=&-T^2\Sigma +T^2(\mu +\Sigma )\log (1+\Sigma /\mu )  \nonumber \\
&&\ +y_c^2(\mu +\Sigma )(\widehat{f}_c\ (b_0)-\widehat{f}_c\ (b_0+\frac{%
2T\Sigma }{y_c\mu (\mu +\Sigma )})  \nonumber \\
&&\ +\frac{2T\Sigma y_c}\mu \widehat{f}_c\ ^{\prime }\ (b_0+\frac{2T\Sigma }{%
y_c\mu (\mu +\Sigma )}).
\end{eqnarray}
These equations possess a RSB solution (they always posses the solution $%
\Sigma =0$) below a temperature $T_c$, which for small $\mu $ is given by 
\begin{equation}
\gamma \widetilde{T}_c=\left( \frac{2g}\pi \right) ^{1/3}\widetilde{B}%
^{-1/6}.
\end{equation}
Just below $T_c$ we find to leading order in $T_c-T$ : 
\begin{eqnarray}
\widetilde{\Sigma } &\sim &\frac 32\widetilde{B}\ (\frac{\widetilde{T}_c-%
\widetilde{T}}{\widetilde{T}_c}), \\
y_c &\sim &\frac 3{\gamma \widetilde{B}^{1/2}}(1-3\frac{\widetilde{T}_c-%
\widetilde{T}}{\widetilde{T}_c}).
\end{eqnarray}

\newpage
Figure Captions:

Fig1: Transverse fluctuations vs. temperature in the cage model for fixed $%
B=250G$. (a) no disorder (b)point disorder ($\widetilde{\Delta }/(2\pi
)=0.2) $ (c)RSB for point disorder, $T<T_{cp}$, (d)columnar disorder ($%
g/(2\pi )=0.025$ )

Fig. 2: Melting line for different amount of point and columnar disorder.
The experimental points from ref. \cite{Khaykovitch} are denoted by symbols
and the theoretical prediction by lines. (a) squares: as grown sample,
continuous curve $\widetilde{\Delta }/(2\pi )=0.144.$ (b) circles: point
disorder induced by a dose of $3\times 10^{18}e^{-}/cm^2$, dashed line $%
\widetilde{\Delta }/(2\pi )=0.208.$ (c) triangles: point disorder induced by
a dose of $6\times 10^{18}e^{-}/cm^2$, dotted line $\widetilde{\Delta }%
/(2\pi )=0.280$. (d) inverted triangles: columnar disorder equivalent to $%
B_\phi =50G$, dashed-double-dotted line $g/(2\pi )=0.01$ and $\widetilde{%
\Delta }/(2\pi )=0.144$. (e) diamonds: columnar disorder equivalent to $%
B_\phi =100G$, dashed-dotted line $g/(2\pi )=0.025$ and $\widetilde{\Delta }%
/(2\pi )=0.144$.


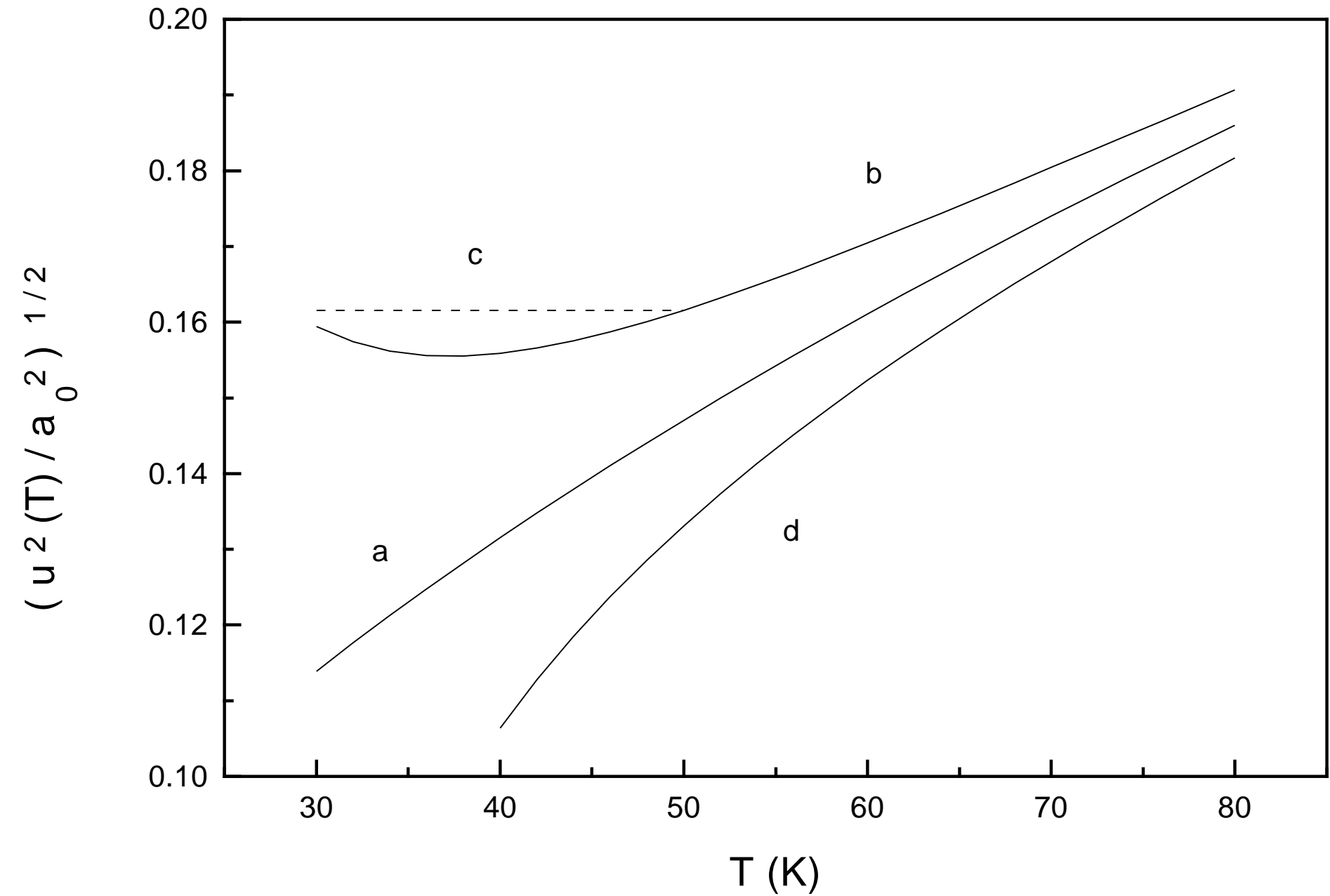
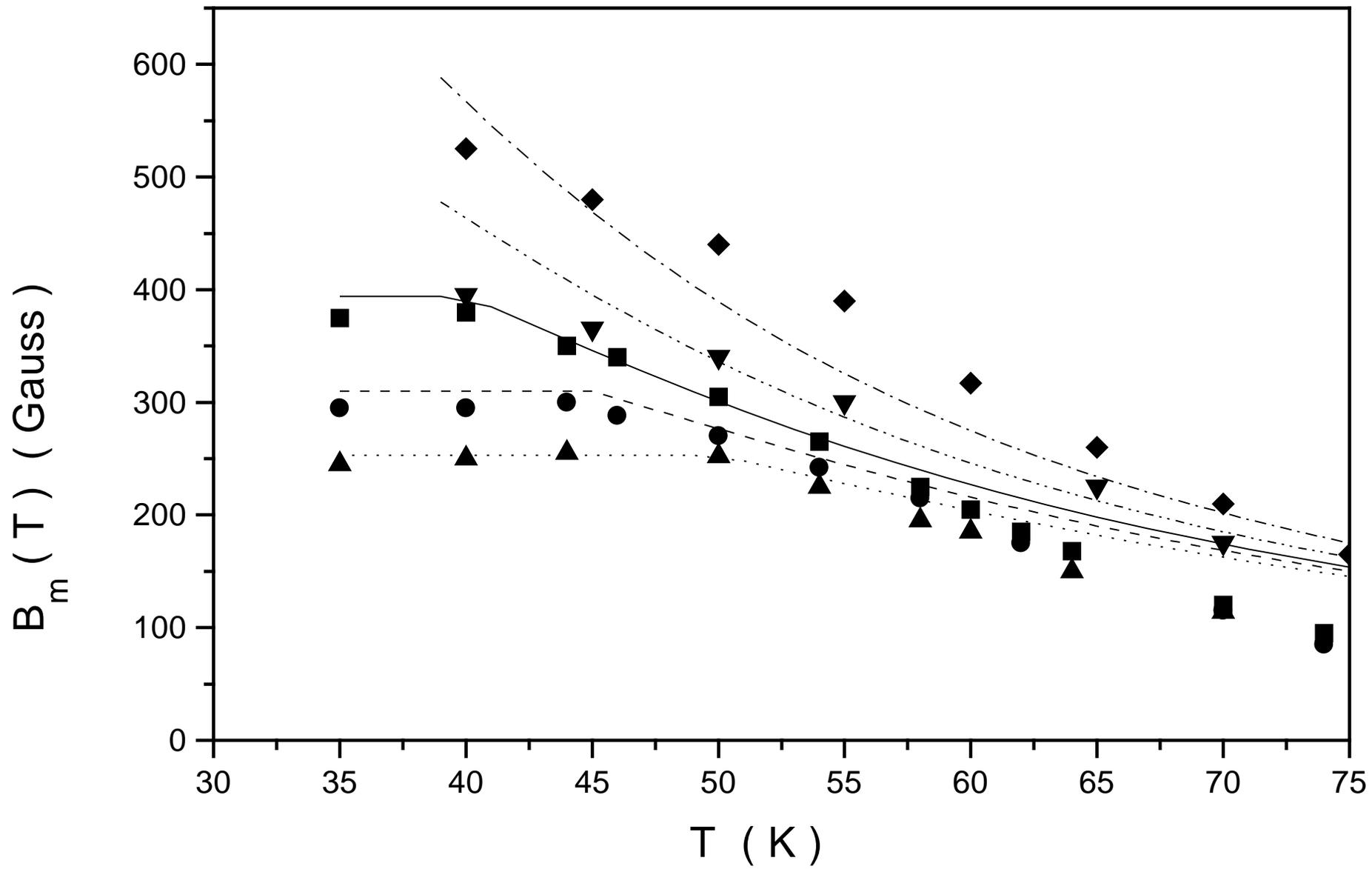
\end{document}